\documentclass[aps,pre,twocolumn,superscriptaddress,showpacs]{revtex4}
\usepackage{epsfig}
\usepackage{amsmath}

\begin{document}

\title{Construction of a secure cryptosystem based on spatiotemporal chaos
and its application in public channel cryptography}
\author{Xingang Wang}
\affiliation{Temasek Laboratories, National University of Singapore, 117508, Singapore}
\author{Meng Zhan}
\affiliation{Temasek Laboratories, National University of Singapore, 117508, Singapore}
\author{Xiaofeng Gong}
\affiliation{Temasek Laboratories, National University of Singapore, 117508, Singapore}
\author{Choy Heng Lai}
\affiliation{Department of Physics, National University of Singapore, 117542, Singapore}

\begin{abstract}
By combining the one-way coupled chaotic map lattice system with a
bit-reverse operation, we construct a new cryptosystem which is extremely
sensitive to the system parameters even for low-dimensional systems. The
security of this new algorithm is investigated and mechanism of the
sensitivity is analyzed. We further apply this cryptosystem to the public
channel cryptography, based on "Merkle's puzzles", by employing it both as
pseudo-random-number (PN) generators and symmetric encryptor. With the
properties of spatiotemporal chaos, the new scheme is rich with new features
and shows some advantages in comparison with the conventional ones.
\end{abstract}

\pacs{05.45.Vx}
\maketitle

\textbf{One serious problem in chaos synchronization based encryptions is
that parameters close to the secret key can still synchronize systems to a
certain extent, and usually appears a regular structure in the key space
under the known-plaintext attack. In other words, there is always a key
basin of finite width around the secret key. The keys in this basin are
highly correlated, and the system security is broken once the location of
the key basin is located and explored. Although in some methods, like those
exploiting spatiotemporal chaos or the modulo operation, can complicate the
cryptanalysis, system security is still vulnerable because the basin width
increases monotonically with the amount of known plaintext. In this paper,
we incorporate a conventional bit operation, the bit-reverse operation, into
spatiotemporal chaos, and find that the basin width shrinks to zero in terms
of the computational precision. This approach not only extends the
definition of the secret key to the real domain and enlarges the capacity of
the key space accordingly, but also overcomes the problem of correlations in
key basin entirely. Using the proposed cryptosystem both as a set of
pseudo-random-number (PN) generators and a symmetric encryptor, we further
investigate the feasibility of public channel cryptography based on chaos,
and propose a prototype for application. In comparison with the conventional
methods, the new model is superior in many aspects: flexibility,
manageability, and simplicity. These new features, together with the
experimental progress in chaos-based communication, make this scheme a good
candidate for public channel cryptography both in software and hardware.}

\section{Introduction}

As an important application of chaos, chaos-based secure communication and
cryptography attracted continuous interest over the last decade \cite%
{pecora,Carroll-IEEE,kocarev,cuomo,algorithm,roy,spatioencryption,ocml-hu-1,
ocml-hu-2,ocml-hu-3,EFA,encoding}. For convenience and flexibility, most of
the proposed schemes are based on the phenomenon of chaos synchronization,
where two chaotic systems can be synchronized through driving or coupling 
\cite{pecora,syn-rep}. While synchronization brings certain advantages for
practical applications, it also presents some drawbacks on the system
security \cite{attacks-1,attack-2,attack-3}. Later it is found that even for high
dimensional chaotic systems, which usually possess higher complexity and
multiple positive Lyapunov exponents, the system security is still
vulnerable under some sophisticated attacks \cite{EFA}. Besides the problem
of security, in comparison to those conventional schemes used widely in
engineering, the performance of chaos encryptions are also disappointing in
other aspects such as having low encryption speed and high bit error rate,
etc \cite{kocarev-ieee,Dachselt, Fraser}. How to design a secure while
efficient cryptosystem has always been a challenge for the chaos
cryptographer.

More recently, the study of applying one-way coupled map lattices (OCML) for
encryption sheds some new light on this research \cite{ocml-hu-1}. One
significant point of this scheme is that two classical numerical operations,
namely integration and modulation, are incorporated into the chaotic
dynamics. With these operations, system security, as well as other
performance indicators, can be greatly improved to a comparable level with
those of the conventional ones, such as DES and AES \cite{ocml-hu-2}. In a
most recent study \cite{ocml-hu-3}, system security is further improved by
adding a S-box, another technique typically used in conventional
encryptions, to the coupled lattices. As a result the capacity of the key
space is further enlarged and the system becomes even more sensitive on the
parameters. However, these schemes still suffer from the problem of the
"continuity" of chaotic dynamics, i.e., correlations still exist between the
keys \cite{EFA}.

In conventional cryptography, encryption schemes are divided into symmetric
and asymmetric methods \cite{bruce-book}. In contrast to the symmetric
methods, the keys in the asymmetric methods are generated in pairs, a public
key and a private key, and it is computationally not feasible to deduce the
private key from the public key. Anyone with the public key can encrypt a
message but not decrypt it. Only the person with the private key can decrypt
the message. Mathematically, the process is based on the trap-door one-way
functions, and encryption is the easy direction and decryption is the
difficult direction. Communication strategies that use asymmetric methods
for encryption have much greater inherent security than symmetric methods,
since they eliminate the problem of key distribution, which itself can pose
the most serious security risk. However, most of the proposed chaos-based
encryption schemes are within the branch of symmetric methods, and little
attention has been paid to asymmetric encryptions, or public-key
cryptography (PKC) \cite{PKC-chaos}. Whereas all known PKC algorithms are
based on some hard problems in number theory (factoriation, knapsack,
discrete logarithms, etc.), it is of great interest and challenge to
construct PKC algorithms based on dynamics.

In the present work, we propose a new scheme of chaos-based symmetric
encryption and, using the proposed cryptosystem both as symmetric encryptor
and pseudo-random-number (PN) generators, design a prototype for public
channel cryptography. In the new cryptosystem, the outputs are extremely
sensitive to the secret key. Any detectable mismatch of the secret key, of
the order of the computer precision, will induces a totally different set of
outputs. Hence this scheme not only overcomes the basic problem of
"continuity" met in chaos-based encryptions, but also extends the definition
of the secret key to all real values in the key space. Borrowing the concept
of "Merkle's Puzzles" \cite{Merkle-puzzle}, we further construct a new model
for public channel cryptography where all blocks are endowed with
spatiotemporal chaos. In comparison with conventional methods, the new model
is found to be more efficient and flexible in some aspects.

This paper is arranged as follows. In Section II we describe our new method
for constructing chaos-based cryptosystems and, in Section III, we give a
detailed discussion of its sensitivity and security. The prototype for PKC
is presented in Section IV, and the system security is analyzed in Section
V. We highlight the new features and advantages of the PKC in Section VI.

\section{Constructing cryptosystem of high security}

As cryptosystems based on low dimensional chaos have been shown \cite%
{attacks-1} to be vulnerable, there have been several efforts to improve the
security by employing spatiotemporal chaos \cite{spatioencryption}. Although
these cryptosystems perform well against some conventional attacks (like the
differential and linear attacks), and can even resist some classical
chaos-based attacks (like the return map and reconstruction attacks \cite%
{attacks-1}), they still suffer some inherent drawbacks from chaos dynamics 
\cite{kocarev-ieee,EFA}. For example, when chaos synchronization is used for
encryption, the keys close to the secret key can still synchronize the
receiver system to a certain extent, thus forming a key basin around the
secret key. (For more details about the definition of key basin, 
please refer Ref. \cite {ocml-hu-1, EFA}.) 
Since the system security is directly connected with the
structure of this basin, it be broken down once the location of this basin
is explored. Based on this, an effective known-plaintext attack \cite%
{bruce-book}, the error function attack (EFA), has been proposed
specifically for cracking chaos synchronization based cryptosystems \cite%
{EFA}. It is found that, under EFA, most of the proposed cryptosystems are
vulnerable or not secure at all, and for some situations the higher
dimensionality does not help to improve system security.

The underlying reason for this "continuity" is that the Lyapunov exponent
(LE) in conventional chaotic systems is not large enough to quickly diffuse
the nearby states in phase space. It is thus natural to look to the
exploration and construction of chaotic systems with large LE for chaos
cryptography, at least as far as EFA attack is concerned. Along this
direction, two methods have been proposed \cite{ocml-hu-2}: (1) using
several of the last significant digits as the output signals and, (2)
coupling lattices with a weak signal. Through these methods, system
sensitivity can be significantly improved, and the width of the key basin
shrinks accordingly. However, as pointed out in Ref. \cite{ocml-hu-3}, there
still exists a scaling between the amount of known plaintext and the width
of the key basin: the more plaintext is known, the wider the key basin will
be. In this respect, the problem of the key basin remains fundamentally
unsolved.

We extend the study in Ref. \cite{ocml-hu-3} and aim to design cryptosystems
that are "truly"' secure. By this we mean cryptosystems with the property
that the sizes of the key basins are of the order of the computational
precision (or the measure precision in practice), and which remain
unchanged with the amount of plaintext known to the attacker. Instead of the
S-box, we construct the transmitter by incorporating a bit-reverse
operation, $F$, into the one-way lattice ring of $N$ coupled logistic maps,
and the dynamics of the transmitter can be formulated as

\begin{eqnarray}
x_{0}(n) &=&S_{N}(n)/2^{\upsilon },  \notag \\
x_{1}(n+1) &=&(1-\varepsilon _{1})f[x_{1}(n)]+\varepsilon _{1}f[x_{0}(n)], 
\notag \\
x_{2}(n+1) &=&(1-\varepsilon _{2})f[x_{2}(n)]+\varepsilon
_{2}f\{F[x_{1}(n)]/2^{\upsilon }\},  \label{trans} \\
x_{i}(n+1) &=&(1-\varepsilon _{i})f[x_{i}(n)]+\varepsilon _{i}f[x_{i-1}(n)],
\notag \\
f &=&4x(1-x),\text{ \ }i=3,4,...,N,  \notag
\end{eqnarray}%
with 
\begin{eqnarray}
S_{N}(n) &=&\{\text{int}[x_{N}(n)\times 10^{h}]\}\text{ mod }2^{\upsilon }, 
\notag \\
F(x) &=&\text{Reverse}\{\text{int}[x\times 10^{h}]\text{ mod }2^{\upsilon
}\}.  \label{reverse}
\end{eqnarray}%
Reverse$\{$ $\}$ represents a bit-reverse operation which reverses the bit
string of an integer and generate another integer as the output. $%
2^{\upsilon }$ is a large integer and $10^{-h}$ is the computer precision.

The dynamics of the receiver (denoted by variables $y_{i}(n)$, $i=1,2,...,N$%
) is identical to that of the transmitter except that the first lattice, $%
y_{1}(n)$, is driven by $x_{0}(n)$. It can be proved that the two systems
can be synchronized under the same driver signal, $x_{0}(n)$, given $%
\varepsilon _{i}>0.75$, $i=1,2,...,N$. In our model, we fix $%
\varepsilon_{i}=0.95$, $i=2,...,N$, and adopt $\varepsilon _{1}$ as the
secret key and define the key space as $\varepsilon _{1}\in \lbrack 0.95,1)$.

For encryption, at the transmitter side, each lattice except the first one
can be regarded as an encryptor. To encrypt a message $P_{i}(n)$ in the $i$%
th channel, we simply perform an XOR (exclusive OR) operation on this
message with the last significant $\upsilon $ bits of the information of $%
x_{i}(n)$, and the output ciphertext reads%
\begin{eqnarray}
C_{i}(n) &=&\text{XOR }[P_{i}(n),X_{i}(n)],  \notag \\
X_{i}(n) &=&\{\text{int}[x_{i}(n)\times 10^{h}]\}\text{ mod }2^{\upsilon },%
\text{ \ }i=2,...,N  \label{integer}
\end{eqnarray}%
The ciphertexts, $C_{i}(n)$, and driver signal $x_{0}(n)$ are then
transmitted to the receiver. The receiver recovers the transmitted message
through the function%
\begin{equation}
P_{i}^{\prime }(n)=\text{XOR }[C_{i}(n),Y_{i}(n)],\text{ \ }i=2,...,N
\end{equation}%
with $Y_{i}(n)$ having the same definition as $X_{i}(n)$ but at the receiver
end. With the same secret key, $\varepsilon _{1}$, the two systems, $x$ and $%
y$, can be completely synchronized, and we finally have $P_{i}^{\prime
}(n)=P_{i}(n)$.

\section{Security analysis}

The key point of this cryptosystem is the bit-reverse operation adopted in
Eqs. \ref{trans}. Since the only secret of symmetric encryption is the key,
the central task of such a cryptosystem is to make the outputs, $X_{i}$, as
sensitive to the secret key as possible. In this scheme, any detectable
mismatch of $\varepsilon _{1}$ (of the order of computer precision) will
affect at least the value of the last bit in $X_{1}$. Due to the bit-reverse
function, this last significant bit becomes the most significant one when
coupled to $x_{2}$, and thus induces a large difference in $X_{2}$ and other
outputs as well. This is further reflected in the behavior of the LE: the
bit-reverse operation is equal to increasing the largest LE (LLE) with a
value of about $h\ln 10$. Thus, the LLE in the newly constructed cryptosystem is
estimated to be%
\begin{equation}
\lambda ^{\prime }\approx \lambda +h\ln 10,
\end{equation}%
with $\lambda $ being the original LLE\ of OCML. For computations with
double arithmetic precision, $h=16$, and with the last $\upsilon =30$ bits of
information adopted as the outputs, the value of the LLE\ for $N=5$ coupled
lattices is about $\lambda ^{\prime }\approx 45$, a value which can diffuse
any detectable mismatch of the secret key, $\varepsilon _{1}$, to the order
of its key space within a few iterations, and thus totally confuses the
"continuity" property in chaos dynamics.

For an eavesdropper, it is easier to attack the $2$nd channel than the
others (studies show that the security of the encryption channel increases
exponentially with the size of the OCML \cite{ocml-hu-2}). We will thus focus on
evaluating the security of this channel in the following. Assuming that the
eavesdropper knows the whole dynamics of Eqs. \ref{trans} and can find an
large amount of plaintext-ciphertext pairs, all he/she needs is to explore
the secret key, $\varepsilon _{1}$, or the key basin where it is located
(we consider here the most common attack used in cryptanalysis: the
known-plaintext attack). By trying some test keys, $\varepsilon _{1}^{\prime
}$, the eavesdropper can study the structure of the key basin by the EFA
function \cite{EFA},%
\begin{equation}
e_{2}(\varepsilon _{1}^{\prime })=\frac{1}{T}\overset{T}{\underset{n=1}{\sum 
}}\left\vert P_{2,\varepsilon _{1}^{\prime }}^{\prime }(n)-P_{2,\varepsilon
_{1}}(n)\right\vert ,
\end{equation}%
with $T$ the amount of known plaintext and $P_{2,\varepsilon _{1}^{\prime
}}^{\prime }$ is the test plaintext generated under the test key $\varepsilon _{1}^{\prime }$%
. Usually there will exist a key basin of a certain width around the secret
key, and the system security will be compromised once this basin is explored.

\begin{figure}[tbp]
\epsfig{figure=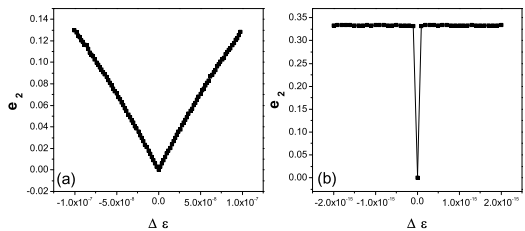,width=0.8\linewidth}
\caption{For OCMLs of size $N=5$ and $T=2 \times 10^{6}$ known plaintexts,
the EFA results of the second channel for encryption schemes: (a) proposed
in Ref. \protect\cite{ocml-hu-1}, and (b) proposed in Section II. The width
of the key basin in (b)\ is the same as the computer precision, $10^{-16}$,
and does not change as $T$ increases.}
\label{fig:EFA}
\end{figure}

In Fig. 1(a), we plot the EFA result of the model used in Ref. \cite%
{ocml-hu-1} with respect to the mismatch between the test key and the secret
key, $\Delta \varepsilon =\varepsilon _{1}^{\prime }-\varepsilon _{1}$. It
can be found, around the secret key, that there exists a smooth basin at
least with a width of $10^{-7}$. With this basin structure, once the
location of the key basin be explored, one can easily get close to the
secret key, which is located at the bottom of the key basin, using only
several test keys by some optimized searching methods. As a comparison, we also plot
in Fig. 1(b) the EFA result of Eqs. \ref{trans}. It is found that the width
of the key basin is just the same as the computer precision $10^{-16}$. The
interesting feature is that, in Fig. 1(b), the width of the key basin does
not increase with $T$. We plot Figs. 1(a) and (b)\ using $T=2\times 10^{6}$
known plaintexts, and had also tested different values of $T$ up to $10^{9}$. The
results confirmed that there is no change for these structures, and that the
basin in Fig. 1(b)\ still has the width of the order of the computer
precision. This property can be immensely useful in preventing attempts to
undermine the system security by studying the key basin structure (according
to the study of Ref. \cite{ocml-hu-3}, even in systems where the modulo
operation is adopted, the relation between the key basin width, $W$, and the
amount of known plaintext, $T$, follows the scaling $W\propto T^{0.3}$).

Two points make this new cryptosystem distinctive and advantage to other
schemes. Firstly, the capacity of the key space can be further extended.
Every real value in the key space can be regarded as an independent secret
key, and the number of independent keys in the key space is limited only by
the computer precision. Secondly, the inherent property of "continuity" in
chaotic systems is now avoided entirely at the level of computational
precision. This renders it hopeless for those attacks based on analyzing the
structure of the key basin. For other encryption performance indicators
(such as the properties of diffusion and confusion, correlations,
robustness, etc.), our numerical simulations \cite{spread-spectrum-STC}
confirmed that there is no difference between this new cryptosystem and the
former schemes (Ref. \cite{ocml-hu-1,ocml-hu-2}).

\section{Applying chaos-based cryptosystem for public channel cryptography}

Besides encryption, due to its excellence performance on statistical
properties, the proposed cryptosystem also can be used as a set of
pseudo-random-number (PN) generators. For this purpose, each lattice can be
regarded as an independent PN generator, and all these generators produce PN
sequences simultaneously. We have checked the random properties of these
sequences with different types of evaluations (such as the run distribution,
balance, power spectrum density, etc.) for arbitrary plaintexts, and they
passed all these checkings satisfactorily \cite{spread-spectrum-STC}. In
addition, in comparison with the conventional PN sequences, these new
sequences possess extremely long periods which increase exponentially both
with the system size and the computer precision. Another interesting observation is that although
there is no statistical correlation between these sequences, teh lattices are
still under the dynamical relation of generalized synchronization (GS) \cite%
{GS}. This special property can be of great use in certain situations where
a large number of independent PN generators are required to operate simultaneously, 
and yet are to be kept in step in some sense. 
The GS relation between lattices also makes it possible to
manipulate all these generators with only a few controllers. Rather than adjusting all
parameters in the generators, now we are able to generate a totally
different set of PN sequences through resetting only one or a few parameters.

In the field of conventional cryptography, there is one type of PKC, namely
the "Merkle's Puzzles", whose security depends on the protocol rather than
number theory. Different to the other PKC schemes, where both the public key
and private key are predefined, in "Merkle's Puzzles", both keys are decided
by the receiver at random, and the keys will be destroyed after each
transmission. A set of independent PN generators and one efficient symmetric
encryptor are the basic blocks for this PKC.  In conventional methods,
usually it is difficult to manage (mainly store and compare) such a large number
of PNs; it is also not easy to find a symmetric encryptor whose security
can be adjusted flexibly so as to keep pace with the improving computer speed.
In this section, we will apply the above proposed cryptosystem on "Merkle's
Puzzles".

\begin{figure}[tbp]
\epsfig{figure=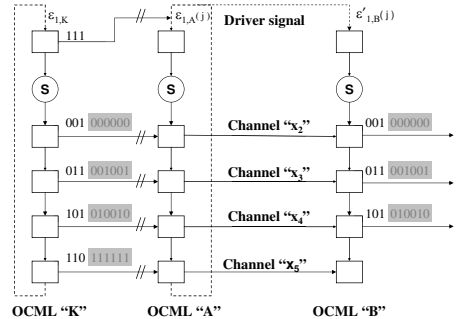,width=0.8\linewidth}
\caption{Prototype for public channel cryptography constructed by three
OCMLs. The dash lines represent feedback or driving signals, shadowed numbers represent
the identifying codes, and "//" means OCML "K" is triggered each time $%
T^{\prime }$.}
\label{fig:PKC}
\end{figure}

The prototype of the PKC is plotted in Fig. 2. The transmitter is composed of
two OCMLs, OCML "K" ("K") used as PN generators and OCML "A" ("A") used as
symmetric encryptor. The receiver comprises the decryptor OCML "B" ("B").
All OCMLs follow Eqs. \ref{trans}. Without "K", the dynamics of the
transmitter is identical to that of the receiver, and it is just the
cryptosystem for symmetric encryption proposed in Section II. "S" represents
the bit-reverse operation in Eqs. \ref{reverse}. "K" has two functions: (1)
generating plaintext for "A" and, (2) modulating the coupling strength of
the first lattice (which is used as the session key for the symmetric
encryptions between "A" and "B"), $\varepsilon _{1,A}$, in "A". "K" is triggered for each time
interval $T^{\prime }$, a session during which both the plaintext and $%
\varepsilon _{1,A}$ remain constant.  Following that, in the next $T^{\prime }$
iterations, "A" encrypts the plaintext outputted from "K" repeatedly under the
session key $\varepsilon _{1,A}(j)$, with $j$ the iteration time of "K".

For the transmitter, the only secret is the parameter $\varepsilon_{1,K}$.
Both the dynamics, "K" and "A", and the initial conditions of "K" are
public. The transmitter has two missions: producing a large number of
encryption sessions and deducing the private key chosen by the receiver. For
the receiver, the dynamics is public and, before deciding on the public keys,
the authorized receiver has no privilege over the eavesdropper. The task of
the receiver is to decrypt one of the transmitted sessions at random, and
returns the decrypted plaintexts - the public keys - to the transmitter
through the public channel.

The details about how to transmit a private key through the public channel can
be described as follows (for OCMLs of size $N=5$):

\begin{enumerate}
\item "K" generates $5$ integers, $X_{i,K}(j)$, $i=1,...,5$, by Eqs. \ref%
{integer}, and marks each of the later four integers with an identification code $%
I_{i,K}$. For instance, in Fig. 2, let us assume the binary format of the
generated integer by $x_{2,K}$ is $X_{2,K}(j)=$`$001$', and mark it with an
identification code $I_{2,K}=$`$000000$'. (For simplicity, the word lengths of the integer
and the identification code here are just used to illustrate the operations, and in
actual simulations both are with the word length of $\upsilon =30$.) The identification
code is only used for marking the channels and is also public. There is no
identification code for $X_{1,K}(j)$, which will be used to modulate the parameter 
$\varepsilon _{1,A}$ in "A". After this, "K" will be dormant until triggered
again for the next session after time $T^{\prime }$.

\item Treating all the marked integers as plaintext, each channel of "A",
according to Eqs. \ref{integer}, encrypts the same plaintext repeatedly for $%
T^{\prime}$ times under the same session key, $\varepsilon _{1,A}(j)$, which
is modulated by the integer $X_{1,K}(j)$ through function%
\begin{equation}
\varepsilon _{1,A}(j)=0.95+\frac{1}{20}X_{1,K}(j)/2^{\upsilon }.
\label{modu}
\end{equation}

\item The transmitter repeats steps (1)\ and (2) until a large number, $L$,
of sessions are generated and transmitted to the receiver.

\item "B" chooses one session at random and performs a brute-force attack to
recover the session key $\varepsilon _{1,A}(j)$ by checking the decrypted
channel identification codes (which are predefined and public) through
synchronization. ($T^{\prime }$ is set so as to ensure that "A" and "B"
can be synchronized for any random initial conditions. In this prototype, $%
T^{\prime }=100$ is large enough for this purpose.) This is a large, but still
manageable, amount of work.

\item After being able to crack one of the sessions successfully, "B" keeps the last
recovered plaintext $X_{5,K}(j)$ as the private key and returns all other
recovered plaintexts, $X_{i,K}(j)$, $i=2,3,4$, to the transmitter together
with their identification codes. The return messages are transferred to "K" in the
form of plaintext and are public to everyone. These plaintexts make up the
set of public keys.

\item After receiving the public keys, the transmitter runs "K" with the
predefined initial conditions (which is also public) and his secret key $%
\varepsilon _{1,K}$ (known only to the transmitter). Once the outputs of the
lattices match up the returned public keys in each corresponding channel
simultaneously, the transmitter will know that the output of the last
lattice, $X_{5,K}(j)$, is the private key which the receiver had chosen, and
which will be used for later communications.
\end{enumerate}

\section{Security of public channel cryptography}

The security of this PKC depends on the number of sessions transmitted. The
eavesdropper can break this system, but he has to do far more work than
either the transmitter or the receiver. To recover the private key $%
X_{5,K}(j)$ in steps (4) and (5), on average, he has to perform a brute-force attack
against about half of the transmitted sessions generated in step (3). Assuming that in total
there are $L$ sessions transmitted in the public channel, the attack of the
eavesdropper has a complexity of $L/2$ times that of the receiver. The public
keys, $X_{i,K}(j)$, $i=2,3,4$, will not help the eavesdropper either; they
are independent PNs generated by the cryptosystem Eqs. \ref{trans}. In
general, the eavesdropper has to expend approximately the square of the
effort that the receiver expends. This advantage is small by cryptographic
standards, but in some circumstances it may be enough. For instance, in
simulations (on a Pentium computer of 2GHZ CPU and 521M RAM, Fortran90 compiler), we set
the duration for each session as $T^{\prime }=100$ and the key space of the
range $\varepsilon _{1,A}\in \lbrack 0.95,0.95+1\times 10^{-8}]$, the
transmitter can generate about $L\approx 1\times 10^{8}$ sessions in one
minute, and the receiver needs another minute to explore one session key $%
\varepsilon _{1,A}(j)$. However, with the same computing facilities, it will
take the eavesdropper about two years to break the system, a time that is
likely to be longer than the useful lifetime of the secret message.

The eavesdropper can of course attack only the private key $X_{5,K}$ used in
the later communications, without considering the problem of PKC. But with
the system under consideration, the private key can be combined randomly and
adjusted freely both in length and position. While this add no additional
cost to PKC, it will be a disaster for an eavesdropper and he/she finally
has to fall back on attacking the sessions. Meanwhile, the excellent
performance on correlations of the system prevents any attempt to deduce the
private key $X_{5,K}(j)$ from the public keys $X_{i,K}(j)$, $i=2,3,4$. The
knowledge of the initial conditions cannot help with predicting $\varepsilon
_{1,K}(j)$ or $X_{5,K}(j)$ either. With the bit-reverse operation, the
difference between two corresponding outputs, $\Delta X_{i,K}(j)=\left\vert
X_{i,K}(j)-X_{i,K^{\prime }}(j)\right\vert $, increases to the order of
attractor size within a few iterations, and after that the behavior of the
two systems are totally different. (For example, with $N=5$ and $\Delta
\varepsilon =\varepsilon _{1,K}-\varepsilon _{1,K^{\prime }}=10^{-16}$, it
needs only about $5$ iterations on average for $\Delta X>1/3$, a smiple criterion in testing randomness \cite{ocml-hu-1}.) So the only
thing the eavesdropper can do is to find out the secret key $\varepsilon
_{1,K}$. The problem of security returns to that of the symmetric
cryptography and, according to our security analysis in Section III, there
is no shortcut but to try all the possible values of $\varepsilon _{1,K}$ in 
$[0.95,1)$ or an even larger range.

In summary, the practical security of PKC only relies on that of the
symmetric encryption, both for the PN generators, "K", and the encryptor,
"A". Given that there is no systematic cryptanalysis developed for the new
cryptosystem, the proposed PKC will be secure.

\section{Discussion and conclusion}

While the incorporated bit-reverse operation improves the security of the
chaos-based cryptosystem to a new level, the adaptation of this cryptosystem
for PKC brings new features and advantages for other real applications as
well.

\begin{itemize}
\item Unlike conventional approaches, the same cryptosystem, Eqs. \ref{trans}%
, can be used both as encryptor and PN generators. This feature can bring
certain convenience both for security analysis and model design.

\item In conventional methods, the transmitter has to store all the PNs in a
group and find the private key which matches up the returned public keys by
a brute-force comparison, which usually involve large amounts of memory
space and computer resource. By adopting "K" as the PN generators, all
these keys can be automatically regenerated through the dynamics of OCML.
Since the security of PKC relies on the number of sessions transmitted, this
property also makes it possible to implement PKC in situations where memory
space is scarce and computer speed is limited.

\item Although one could replace each lattice in "K" with a separate
conventional PN generator, in real applications it is usually hard to keep
them working in step. But this problem does not appear for OCML, where all
sequences are outputted simultaneously under the relation of GS.

\item The process of recovering the private key $X_{5,K}(j)$ from the public
keys $X_{i,K}(j)$, $i=2,3,4$, is achieved by the trap-door $\varepsilon
_{1,K}$, the only secret of the transmitter. With the trap-door, it is easy
to recover all keys of the chosen session, but this fails for any detectable
mismatch. In this regard, the proposed OCML actually can be used as a
one-way function with the trap-door $\varepsilon _{1,K}$.
\end{itemize}

The proposed PKC also enjoys all advantages of traditional chaotic systems.
The security of encryptor "A" can be updated easily either by enlarging its
key space or combining more couplings as the session key, which make this
scheme easily adjustable to different security requirements. In addition to
the implementations on software, the proposed scheme is expected to be
efficient on hardware as well, judging from the progress in chaos
experiments \cite{chaosexperiment}. The dynamics based cryptography makes it
not only easy to formulate and analyze system security in theory, but also
simple to design and operate the constructed cryptosystems in applications.
Meanwhile, the performance of PKC can be further enhanced by chaos-based
spread-spectrum communications \cite{spread-spectrum-STC}. Whereas the
security of PKC relies on the number of transmitted sessions, it is highly
recommended to transmit these data through a wide-band channel so as to
achieve a fast speed, and chaotic signals, with their excellent performance
on correlations, can be used for this purpose directly.

In conclusion, we have proposed in this paper a way of improving the security of chaos-based
cryptosystem to the order of measure precision, and applied it to PKC by
using the system both as PN generators and symmetric encryptor.
Incorporating the conventional bit-reverse operation, we successfully
overcome the problem of "continuity" in chaotic systems, and equip the
conventional scheme of PKC with new characteristics of spatiotemporal chaos.

\end{document}